# BCI-Based Strategies on Stroke Rehabilitation with Avatar and FES Feedback

Zhaoyang Qiu[1], Shugeng Chen[2], Brendan Z Allison[3], Jie Jia[2*], Xingyu Wang[1], Jing Jin[1*]

[1]Key Laboratory of Advanced Control and Optimization for Chemical Processes, Ministry of Education, East China University of Science and Technology, Shanghai, P.R. China

[2]Department of Neurology, Huashan hospital, Fudan University, Shanghai, P.R. China

[3]Department of Cognitive Science, University of California at San Diego, La Jolla, California, USA

* jinjingat@gmail.com, shannonjj@126.com

**Abstract**

**Background:** Stroke is the leading cause of serious and long-term disability worldwide. Survivors may recover some motor function after rehabilitation therapy. Some studies have shown that motor imagery (MI) based BCI has a positive effect in poststroke rehabilitation. It could help patients promote the reorganization processes in the damaged brain regions. However, offline motor imagery and conventional online motor imagery with feedback (such as rewarding sounds and movements of an avatar) could not reflect the true intention of the patients.

**Method:** In this study, both virtual limbs and functional electrical stimulation (FES) were used as feedback to provide patients a closed-loop sensorimotor integration for motor rehabilitation. Motor imagery based BCI system acquired, analyzed and recognized the EEG data of motor imagery from patients. The FES system would activate if the user was imagining hand movement of instructed side. Ten stroke patients (7 male, aged 22–70 years, mean 49.5±15.1) were involved in this study. All of them participated in BCI-FES rehabilitation training for 4 weeks.

**Results:** The average motor imagery accuracies of the ten patients in the last week were 71.3%, which has improved 3% than that in the first week. Five patients' Fugl-Meyer Assessment (FMA) scores have been raised. Patient 6, who has have suffered from stroke over two years, achieved the greatest improvement after rehabilitation training (pre FMA: 20, post FMA: 35). In the aspect of brain patterns, the active patterns of the five patients gradually became centralized and shifted to sensorimotor areas (channel C3 and C4) and premotor area (channel FC3 and FC4).

**Conclusions:** In this study, motor imagery based BCI and FES system were combined to provided stoke patients with a closed-loop sensorimotor integration for motor rehabilitation. Result showed evidences that the BCI-FES system is effective in restoring upper extremities motor function in stroke. In future work, more cases are needed to demonstrate its superiority over conventional therapy and explore the potential role of MI in poststroke rehabilitation.

## Introduction

Stroke is one of the most common cerebrovascular diseases worldwide. It is the leading cause of serious and long-term disability in many countries [1-3]. Motor disorders are a frequent consequence of stroke, often including hemiplegia of the upper limbs. After rehabilitation therapy, some stroke survivors can partially regain their motor control [4]. Timely and effective treatment were useful for motor function recovery, which could help the survivors perform daily activities better.

Different rehabilitative approaches have been used for poststroke treatment [5-6]. One of them is Motor Imagery (MI). Motor imagery was one kind of mental practice, which has been considered as an effective neurorehabilitation technique to enhance poststroke motor recovery [7-10]. Braun et al., 2006 conducted a systematic review of all randomized controlled trials that analyze the effect of MI on patients after a cortical stroke [11]. Results showed that motor imagery provides additional benefits to conventional physiotherapy or occupational therapy after stroke. Sjoerd et al., 2007 reviewed the evidence for motor imagery or observation as novel methods in stroke rehabilitation [5]. The literature showed that imagery training may be valuable new methods for post-stroke motor rehabilitation. Floriana et al., 2015 found that better functional outcome was observed in the MI training group, including a significantly higher probability of achieving a clinically relevant increase in the FMA score compared to the control group [12].

Brain-computer interface systems (BCIs) can translate brain activities into commands that could be used to control external devices [13-15]. BCIs provide a new communication channel for people, which do not rely on the conventional neuromuscular pathways of peripheral nerves and muscles [16]. Motor imagery could elicits Event-related (de)synchronization (ERD/ERS) [17, 18] that represents the result of conscious access to the content of the intention of a movement. Therefore, the ERD/ERS feature could be used to detect the motor intention of stroke patients. MI-based BCI systems has increased the number of potential BCI users exponentially, which could support motor rehabilitation. Many groups have tested the applicability of MI based BCIs in stroke rehabilitation [10, 19-22].

However, during the conventional rehabilitation process, there is no objective way to determine whether the patients are performing the expected motor imagery task [23]. In the absence of guidance and proper training, patients usually cannot perform the task well. To help patients modulate brain activity proficiently, many training methods with feedback were used to improve the performance of MI-based BCIs [12, 22]. These feedback controlled by the system, such as rewarding sounds and real-time bar, could not reflect the true intention of the patients. To provide patients a closed-loop sensorimotor integration for motor rehabilitation, functional electrical stimulation (FES) were used as feedback in some studies gradually. Functional electrical stimulation (FES) use a certain intensity pulse current to stimulate one or more groups of muscles. It could induce muscle movement to improve or restore muscle function. In [], a rehabilitation system using FES as feedback was investigated. Results showed that affected motor related cortex of patient subject was activated significantly. In [], researchers used a FES system and an bar as the feedback. After 10 sessions training, one stroke patient partially regained control of dorsiflexion in her paretic wrist.

However, there were too few cases of these studies. [] included 2 stroke patients and only one stroke patient was involved in []. In addition, single pathway feedback is the main form in existing research. In this paper, cues were played by audio and both virtual limbs and functional electrical stimulation (FES) were used as feedback. The avatar of patients' upper limbs provide a virtual reality feedback, which could help patients perform MI tasks better. The FES system use a certain intensity pulse current to stimulate one or more groups of muscles which could provide an efficient proprioceptive feedback. Ten stroke patients were involved in the study. They imagine the left or the right wrist dorsiflexion during the rehabilitation training.

**Methods**

*A. The BCI-FES system*

In this study, the recoveriX system was used for experiments, which is a BCI-FES system actually. EEG signals were sampled at 256 Hz through a g.USBamp (Guger Technologies, Graz, Austria). The band pass filter was set to 0.1-30 Hz. A 16-channel cap (FC3, FCZ, FC4, C5, C3, C1, CZ, C2, C4, C6, CP3, CP1, CPZ, CP2, CP4 and PZ) following the 10-20 international system was used for signal recording. Data were referenced to electrode REF located over the right mastoid with a forehead ground (GND), shown in Fig. 1. The FES system was controlled through a g.STIMbox. It stimulated appointed muscle group, and the modes were triggered by BCI system.

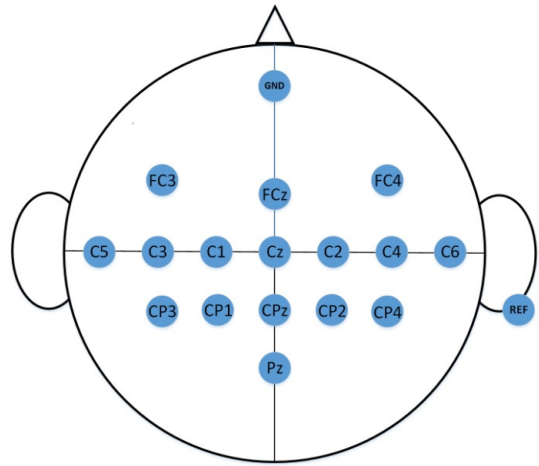

**Fig. 1.** The electrode distribution used in this study.

The schematics of the BCI-FES system was shown in Fig. 2. As stroke patients suffer neurological damage, the brain regions associated with motor function might be compromised and could not control limb movements directly. However, the BCI system could acquire, analyze and recognize the EEG data of motor imagery from patients. The FES system would activate if the user was imagining hand movement of instructed side. The muscle contraction by FES was sufficient enough to cause movement in the affected hand.

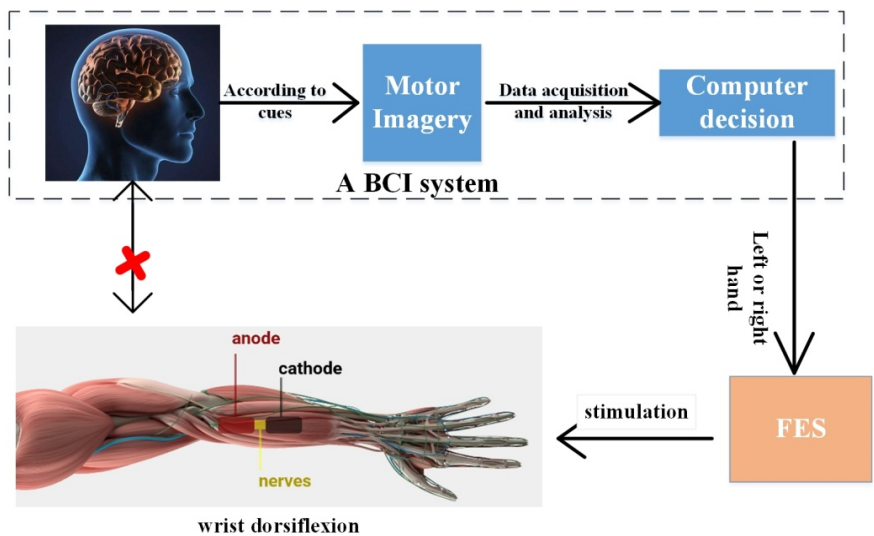

**Fig. 2.** The schematics of the BCI-FES system.

*B. Patients*

Ten stroke patients (7 male, aged 22–70 years, mean 49.5±15.1) participated in this study. Four of them have left hemiplegia, while the rest have right hemiplegia. All patients signed a written consent form prior to this experiment. The local ethics committee approved the consent form and experimental procedure before any patients participated. All patients were right handed according to self-reports. All patients were diagnosed by CT or MRI, without cognitive disorder and any unsuitable diseases for receiving BCI-FES system. Table 1 shows demographic information and motor functions scores for all patients.

TABLE I
INFORMATION OF THE TEN PATIENTS

| ID | Gender | Age | Stroke Type | Paretic Side | Months Since stroke |
|---|---|---|---|---|---|
| 1 | Female | 59 | Trauma | Left | 2 |
| 2 | Female | 70 | Ischemic | Right | 8 |
| 3 | Female | 22 | Trauma | Left | 248 |
| 4 | Male | 65 | Ischemic | Right | 8 |
| 5 | Male | 44 | Hemorrhage | Right | 22 |
| 6 | Male | 45 | Hemorrhage | left | 24 |
| 7 | Male | 30 | Trauma | left | 38 |
| 8 | Male | 56 | Ischemic | Right | 16 |
| 9 | Male | 58 | Hemorrhage | Right | 6 |
| 10 | Male | 46 | Infarction | Right | 2 |

*C. Experimental Procedure*

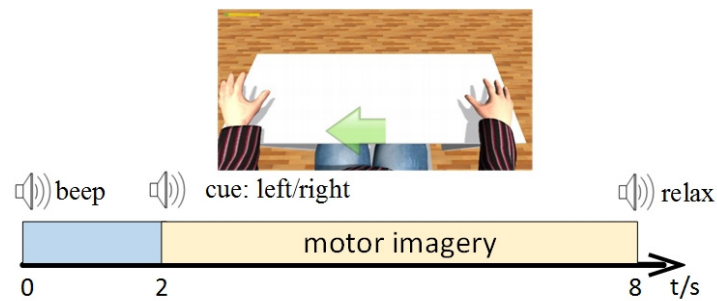

**Fig. 3.** Timing of a trial of the motor imagery paradigm. Each trial consisted of task and rest periods. Patients started to execute motor imagery tasks while the cue ("left" or "right") was played.

The patients were seated in a comfortable chair in a shielded room after being prepared for EEG recording. During data acquisition, patients were asked to relax and avoid unnecessary movement. Instructions were through both sounds and visual cues. The experimenter informed the patients that they would hear cues over a speaker that would instruct them to imagine moving either the left or the right hand. At the same time, the screen would show the hand avatar which reflecting the right movement to help patients imagine the wrist dorsiflexion. Fig. 2 shows that each trial lasts eight seconds and starts with a warning "beep". Two seconds later, the cue (the command to imagine a left or right hand movement) is played. Six second later, a "relax" command is played, informing patients that the trial is over. The FES would activate if the user was imagining hand movement of instructed side. The muscle contraction by FES was sufficient enough to cause movement in the affected hand. The feedback period lasted four seconds, and the inter-trial interval was two seconds. Fig. 4. shows the scene of BCI-FES rehabilitation training for a patient. The avatar of the screen and the FES would give her feedback at the same time when she performed MI task properly. Each patient participated in sixty trials within one recording run. Each session is consisted of sixty trials and lasts for 600s. Three rehabilitation sessions were carried out a week, each containing two runs. Patients participated in BCI-FES rehabilitation training for 12 times in one month (three times per week).

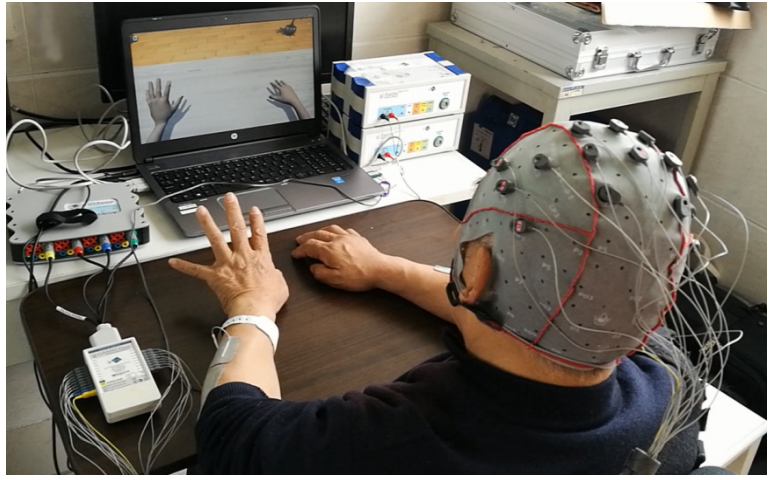

**Fig. 1.** This picture shows the scene of BCI-FES rehabilitation training for Patient 2. The avatar of the screen and the FES would give her feedback at the same time when she performed MI task properly.

*D. Pattern Recognition*

For motor imagery feature extraction, the EEG data were band-pass filtered using a fifth order Butterworth band pass filter from 8 to 30 Hz, since this frequency band included the range of frequencies that were mainly involved in performing motor imagery.

The CSP algorithm is an effective feature extraction algorithm. It has been widely used in processing EEG data from motor imagery [24-29]. CSP is based on the simultaneous diagonalization of two covariance matrices. It finds a spatial filter to maximize variance for one class and minimize variance for another class at the same time to improve classification.

For the analysis, the original EEG signal data is represented as a matrix $E \in R^{N \times T}$, where $E$ is the EEG samples of a trial with dimensions $N \times T$ $N \times T$, $N$ is the number of channels, and $T$ is the number of sampling points for each channel. The CSP operation process is as follows:

Calculate spatial covariance of the EEG data:

$$C = (EE^T)/(tr(EE^T)) \qquad (1)$$

$C_l$ and $C_r$ represent two spatial covariance matrixes (two classes of motor imagery). They can be calculated by averaging over the trials of each group. The composite spatial covariance matrix can be expressed as:

$$C_c = \overline{C_l} + \overline{C_r} \qquad (2)$$

$C_c$ can be decomposed as:

$$C_c = U_c \lambda_c U_c^T \qquad (3)$$

$U_c$ is the matrix of eigenvectors and $\lambda_c$ is the diagonal matrix of eigenvalues. In the process, the eigenvalues are arranged in descending order.

The whitening transformation:

$$P = \sqrt{\lambda_c^{-1}} U_c^T \tag{4}$$

Then the covariance matrix $\overline{C_l}$ and $\overline{C_r}$ can be transformed into:

$$S_l = P\overline{C_l}P^T, S_r = P\overline{C_r}P^T \tag{5}$$

$S_l$ and $S_r$ share the same eigenvectors. If $S_l = B\lambda_r B^T$, then

$$S_r = B\lambda_r B^T, \lambda_l + \lambda_r = I \tag{6}$$

$I$ is the identity matrix. The projection matrix is achieved by the following equation:

$$F = (B^T P)^T \tag{7}$$

This is the expected spatial filter. After whitening, EEG signals can be projected on the first $m$ and last $m$ columns of $B$. So the EEG data of a single trial can be transformed into:

$$Z = FE \tag{8}$$

In this paper, we defined $m = 3$. $f_p$ could be obtained from $Z_p$ ($p = 1 \ldots 2m$) as the features of the original EEG data:

$$f_p = \log\left(\frac{Var(Z_p)}{\sum_{i=1}^{2m} Var(Z_i)}\right) \tag{9}$$

Linear discriminant analysis (LDA) is a generalization of Fisher's linear discriminant, a method used in statistics, pattern recognition and machine learning to find a linear combination of features that characterizes or separates two or more classes of objects or events [30, 31]. The resulting combination may be used as a linear classifier, or, more commonly, for dimensionality reduction before laterclassification.. In this paper,10*10-fold cross validations accuracy of the data were used for each patients.

**Results**

*A. Classification Performance Comparison*

Figure 4 presents the BCI classification accuracy performance across 12 training sessions for the ten patients. Results showed that there were marked upward trends in performance for P2, P6 and P7. The average accuracies of the three patients in the last week were 88%, 77.8% and 73%. The performance have improved a lot than that in the first week ( 76.8%, 71.1% and 60%). The performances of P1 and P9 were relatively stable. P1' s accuracies were basically between 80% and 90% and the accuracies of P9 were almost higher than 95%. P9 achieved the best performance of all the patients. However, the average accuracy of P4 is the worst of all. Since the classifier was attempting to distinguish two classes (right and left hand motor imagery), chance accuracy was 50%. The performance of P9 mean that the FES system could hardly stimulate the right hand, and rehabilitation training was not effective. For the rest patients, the performances fluctuated considerably. According to the patient's report, the decreased accuracies were related to lack of sleep during the previous night and emotional fluctuations.

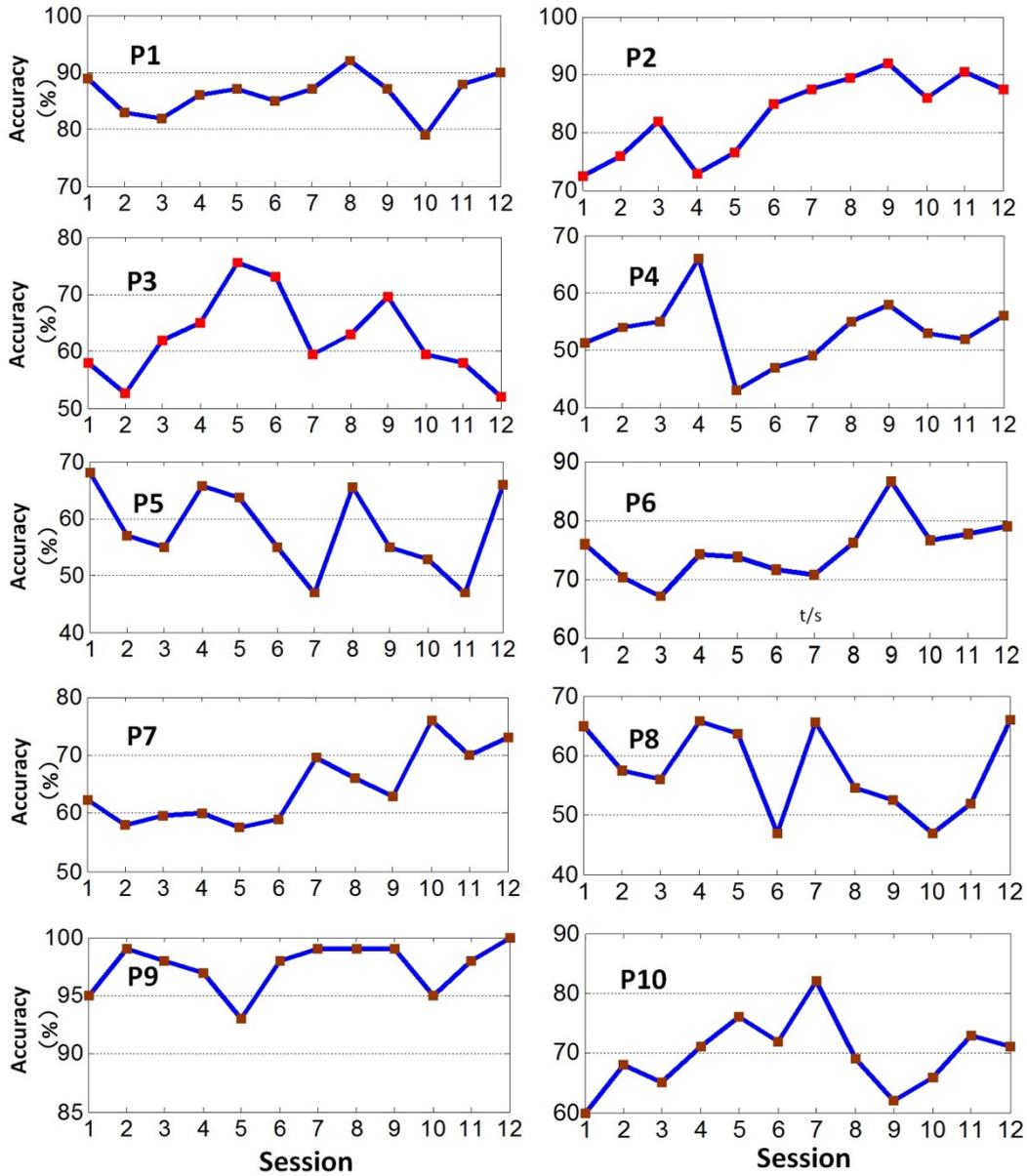

**Fig. 4.** The accuracy plots across 12 sessions for ten patients. "P" represents "patient".

*B. Functional improvement*

Fugl-Meyer Assessment (FMA) scale is an index to assess the sensorimotor impairment in individuals who have had stroke.[1] This scale was first proposed by Axel Fugl-Meyer and his colleagues as a standardized assessment test for post-stroke recovery in their paper titled The post-stroke hemiplegic patient: A method for evaluation of physical performance. It is now widely used for clinical assessment of motor function.[2][3] The Fugl-Meyer Assessment score has been tested several times, and is found to have excellent consistency, responsivity and good accuracy. The FMA assesses several impairment dimensions using a 3-point ordinal scale (0=cannot perform; 1=can= perform partially; 2=can perform fully).

In this paper, FMA was used to evaluate the motor function of upper limb of ten patients. Figure 4 presents the FMA comparison before and after rehabilitation training over the ten patients. Results showed that the score has been raised for P1, P6, P7, P8 and P9. Scores of rest patients did not change. Among all ten patients, P6 achieved the greatest improvement after rehabilitation training(increased by 15). P1 also has made great progress and her score changed from 22 to 32. The increased scores were 4, 6 and 3 for P7, P8 and P9, respectively.

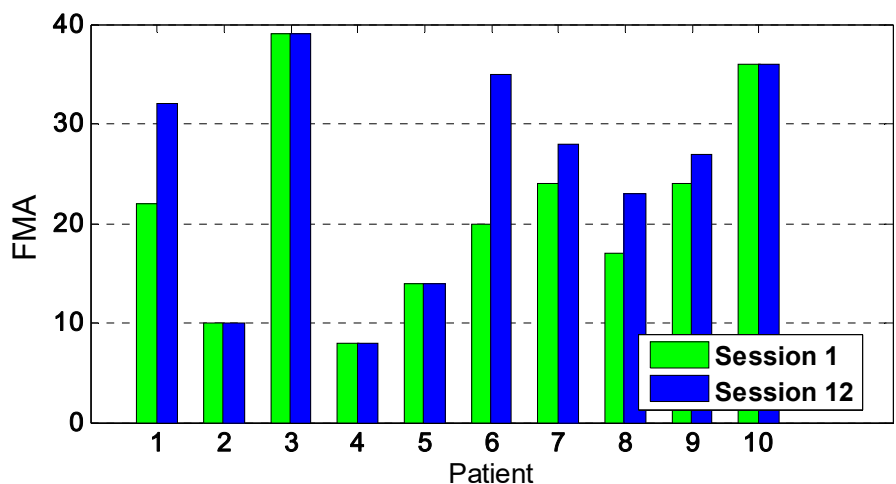

**Fig. 5.** The FMA score comparison before and after rehabilitation training over the ten patients.

*C. Brain patterns*

Fig. 5 showed the power spectral densities (PSDs) calculated from the average data to illustrate the differences between session 1 and session 12 (before and after rehabilitation training using the BCI-FES system). Each map was based on 120 averaged trials (60 left hand motor imagery trials and 60 right hand motor imagery trials) after being filtered by 8-30 Hz band pass filter. In Fig. 5, the blue line represents the left hand motor imagery and the black line represents the right hand motor imagery . For each patient, the figure showed the power spectral densities at C3 and C4 channels, since the two channels have been proved to record important characteristics of motor imagery[].

Fig. 5 presents the PSDs across the five patients whose scores of FMA were improved. For patient 1, the energy of right hand motor imagery was lower than left hand motor imagery at C3 in both session 1 and session 12. The differences mainly occurred in 8-20 Hz. The energy of left hand motor imagery was higher at C4 in session 1 while it was lower than right hand motor imagery in the beta band after rehabilitation training. For patient 6, the energy of left hand motor imagery became lower than right hand motor imagery in the beta band after rehabilitation training. For patient 7, there were not significant differences at C3 and C4 in session 1. However, in session 12, the energy of left hand motor imagery became lower than right hand motor imagery in the beta band. For patient 8, the energy of right hand motor imagery was higher than left hand motor imagery at both C3 and C4. It was hard to distinguish the two motor imagery tasks. After 12 sessions rehabilitation training, the energy of right hand motor imagery became lower than left hand in 16-24 Hz at C3. For patient 9, the PSDs maps were consistent with the theory [] (mainly in the beta band) in both session 1 and session 12: the energy of right hand motor imagery was lower than left hand motor imagery at C3, and the energy of left hand motor imagery was lower than right hand motor imagery at C4.

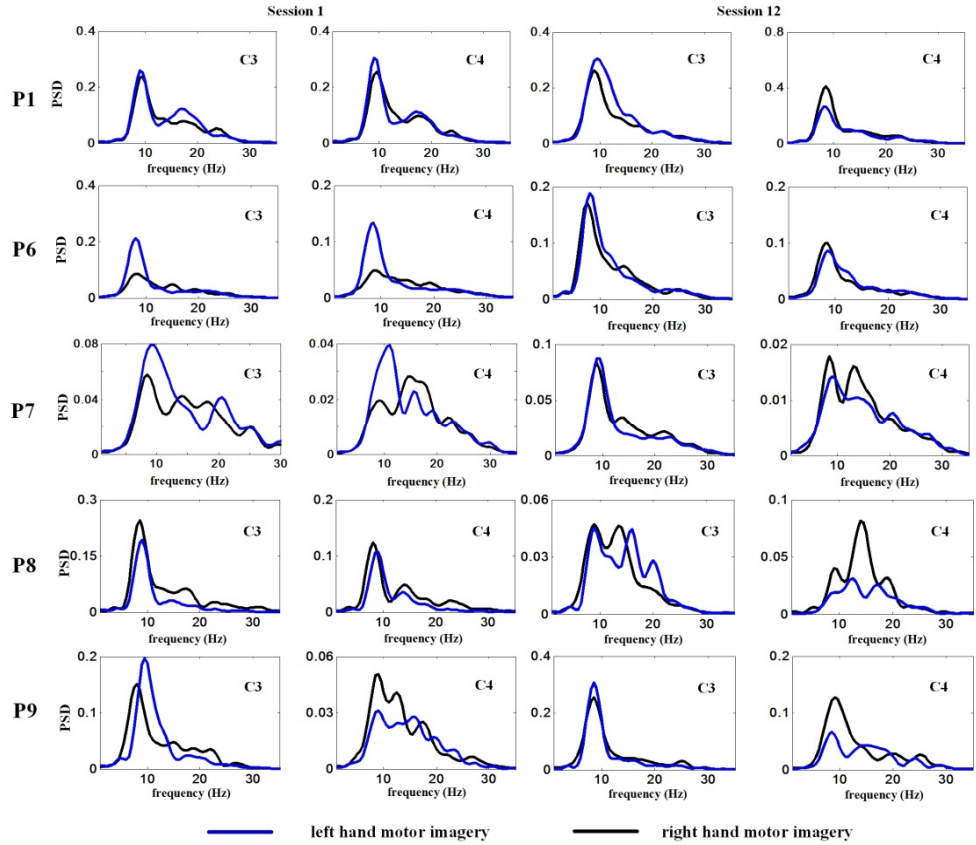

**Fig. 6.** This figure displays the averaged spectra of 120 trials recorded at C3 and C4 for five patients in session 1 and session 12 (blue: right hand, black: left hand). Lines in the map have been smoothed. "P" refers to patient.

In CSP method, W is the projection matrix, and W-1 is the inverse matrix of W. The columns of W-1 are the time invariant vectors of EEG source distribution vectors called common spatial patterns [52]. The first pattern was obtained by maximizing the variance of the right hand motor imagery, which was associated with the ERD phenomenon over the area of the left sensorimotor area of the cortex. Accordingly, the ERD phenomenon over the right motor area was associated with the last pattern, corresponding to the left hand motor imagery. The common spatial patterns of the five patients whose scores of FMA were improved are shown in Fig. 7 . This figure only showed the paretic side of each patient (P1, P6 and P7: left hand motor imagery; P8 and P9: right hand motor imagery) before and after the rehabilitation training.

For patient 1, the ERD phenomenon of left hand motor imagery mainly occurred in the right of central area in the first session. In the last session, the area of ERD phenomenon moved to the right, mainly distributed around channel C4. For patient 6, the area of ERD phenomenon became centralized and mainly distributed around channel C4 after 12 sessions rehabilitation training. In the last session, the ERS phenomenon could be seen in the left cerebral cortex. Results of patient 7 in the first session did not show clear pattern from the contra lateral hemisphere for left hand motor imagery. However, the maps of the last session showed clear ERD phenomenon in the right cerebral cortex. For patient 8, in both the first and last session, the ERD phenomenon were not particularly strong. It mainly occurred in the left of central area for the first session and in the upper left cerebral cortex for the last session. Results of patient 9 showed clear ERD phenomenon in the left cerebral cortex during right hand motor imagery and the locations were different. Fig. 7 also indicated that areas around channel C3 and C4 were strongly associated with the left and right hand motor imageries, which was consistent with the neurophysiology phenomenon reported in [53-54].

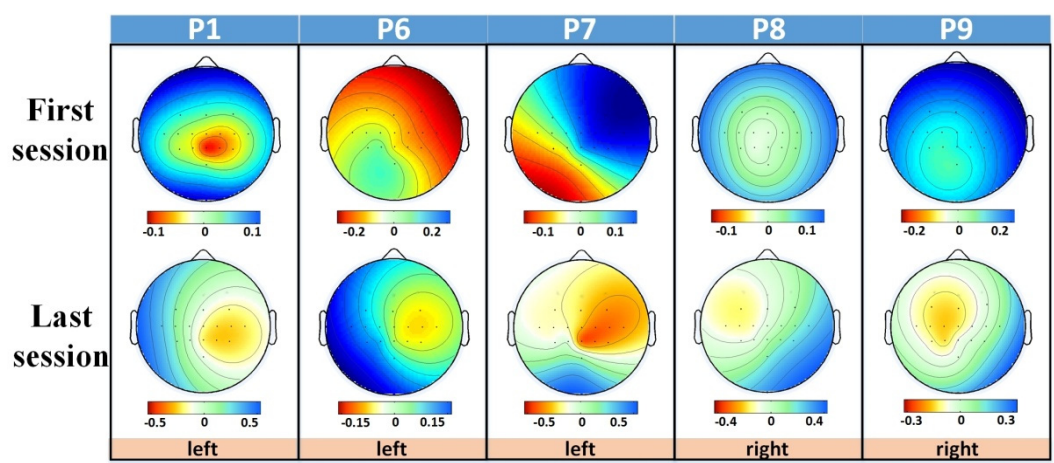

Fig. 7. Ten subjects' topographic maps of the first and last spatial patterns extracted by the CSP method. The patterns were extracted from averages of 100 trials of the two different paradigms.

**4. Discussion**

Different rehabilitative approaches have been used for post stroke rehabilitation. One of them is motor imagery (MI). MI has been adopted in rehabilitation programs for persons with stroke in many researches. To provide patients a closed-loop sensorimotor integration for motor rehabilitation, we combined motor imagery based BCIs and functional electrical stimulation. Both virtual limbs and FES were as feedback in this paper, which could help patients do the rehabilitation training more effectively.

*A. Classification performance*

Figure 4 presents the BCI classification accuracy performance across 12 training sessions for the ten patients. The performances could be divided into three cases roughly: progress, relatively stable and undulating. Results showed that there were marked upward trends in performance for P2, P6 and P7. The performances of P1 and P9 were relatively stable. The performances of P3, P4, P5 and P10 showed relatively large fluctuations. According to the patients' reports, the undulating accuracies were related to lack of sleep during the previous night and emotional fluctuations. After a period of rehabilitation training, some patients found that their capacity of hand action did not improve, they felt frustrated and fidgety in the experiment. This caused them unable to concentrate on motor imagery. Hence, P3, P4, P5 and P10 performed badly in the two category of motor imagery. An interesting observation is that, before this motor imagery based rehabilitation training, P1 and P9 had imagined the hand movements usually according to their reports. Especially for P9, a minor depressed patient, often imagine his body parts. Hence, he achieved the best performance of all the patients and almost all accuracies were above 95%.

In this paper, both virtual limbs and functional electrical stimulation (FES) were as feedback for patients. In many studies, an arrow pointing left or right was used as the cue. However, there was evidence that viewing real or artificial body parts resulted in a stronger desynchronization than viewing non-body part movements [32, 33]. Some researchers recently linked this modulation of sensorimotor brain rhythms in the mu and beta frequency band to the human mirror neuron system (an action observation / execution matching system) [34, 35]. It was an ability of performing an internal simulation of the observed action. Hence, using virtual limbs as feedback could help patients perform the motor imagery tasks better.

*B. Motor function recovery*

Fugl-Meyer Motor Assessment (FMA) (Fugl-Meyer et al., 1975) was used to evaluate the effect of BCI-FES rehabilitation system on motor function recovery. Table 2 showed the motor functions scores before and after the BCI-FES rehabilitation training of the ten patients. Five patients' FMA scores have been increased. Especially for P6, P7 and P8, their sick time were all over 16 months. Their conditions tend to be stable and difficult to progress during the routine rehabilitation training. However, the BCI-FES rehabilitation system reactivated some damaged brain regions, and improved their motor function. On the other hand, there were still five patients who had not obtained any progress. The reason may be that: (1) Patients felt frustrated and fidgety in the experiment and could not concentrate on motor imagery (for P4 and P5). The BCI-FES rehabilitation system was ineffective for them. (2) The sick time was too long and it was difficult to activate the damaged brain regions. For P3, she had suffered from stroke over twenty year, various treatments have little effect. In summary, significant improvements of motor functions of upper limbs has been achieved for some patients. The BCI-FES system provide an effective alternative to traditional treatments on stroke patients' rehabilitation.

TABLE II
TEN SUBJECTS' RESPONSES TO QUESTIONS ABOUT THE TWO PARADIGMS

|  | P1 | P2 | P3 | P4 | P5 | P6 | P7 | P8 | P9 | P10 |
|---|---|---|---|---|---|---|---|---|---|---|
| Time since stroke (months) | 2 | 8 | 248 | 8 | 22 | 24 | 38 | 16 | 6 | 2 |
| Pre FMA (Full = 66) | 22 | 10 | 39 | 8 | 14 | 20 | 24 | 17 | 24 | 36 |
| Post FMA (Full = 66) | 32 | 10 | 39 | 8 | 14 | 35 | 28 | 23 | 27 | 36 |

*C. EEG patterns changes*

In this paper, power spectral densities (PSDs) and topographic maps extracted by the CSP were used to detect motor imagery EEG patterns. By tracking the changes of motor imagery EEG patterns during rehabilitation, we tried to find out the cortex reorganization process. Our results suggest that the sensorimotor cortex in the contralateral hemisphere could be activated after rehabilitation. The sensorimotor area around channel C3 and C4 were strongly associated with the motor imageries, which was important for motor rehabilitation [53-54]. For patient 1 whose paretic side are left, the active areas (ERD) gradually shifted back to channel C4 and gathered around it compared to session 1. For patient 6, larger regions initially take parts of functionality in session 1: the ERD phenomenon almost occurred in the whole contralateral hemisphere. However, the active areas gradually became centralized and mainly distributed around channel C4 after 12 sessions rehabilitation training. For patient 7, there was not ERD phenomenon occurred in the right hemisphere during left hand motor imagery. It might be that the patient has not mastered the skill of motor imagery at start. Hence the topographic map did not show very discriminative patterns. After a month of training, active patterns were produced in the right hemisphere and mainly distributed in motor area and premotor area. For patient 8 and 9, the active areas shifted forward and were produced in premotor area (FC3).

In summary, after 12 sessions rehabilitation training, the active patterns of the five patients gradually became centralized and shifted to sensorimotor areas (around channel C3 and C4) and premotor area (around channel FC3 and FC4). Some literature had reported the similar phenomenon. Tam et al. (2011) reported that the active patterns were produced in premotor areas and the parietal area instead of sensorimotor areas. Liu et al. (2014) reported that parietal lobes (P4) and frontal premotor lobes (FC4) would play a compensatory role but the importance decreases over time. This observation reveals the rehabilitation mechanism: larger regions initially take parts of functionality but gradually given back during recovery[]. It was a sign of cortex reorganization in the affected hemisphere.

Fig. 6 illustrates the power spectral densities of C3 and C4 (filtered from 8 to 30 Hz) for the five patients whose scores of FMA were improved. For patient 1, 6 and 7 in session 1, subtle differences mainly scatter around 8–20 Hz at C4 between the two motor imagery tasks. For patient 8 and 9 in session 1, the differences mainly scatter around 8–25 Hz at C3. However, in session 12, the different bands gradually concentrated to narrower bands for three of the five patients (P1: 8-11 Hz; P6: 8-14 Hz; P7: 9-11Hz). Some researchers had reported that active frequency bands decentralize at wide-ranged bands at the beginning and gradually concentrated to narrower bands. The dynamic migration on frequency bands implies that active rhythms are modulated during rehabilitation []. In addition, the difference between the two motor imagery tasks showed the correct phenomenon after 12 sessions rehabilitation training: the energy of left hand was lower than right hand at C4 during left hand motor imagery, and the energy of right hand motor imagery was lower than left hand at C3.

## 5. Conclusion

In this study, motor imagery based BCI and FES system were combined to provided stoke patients with a closed-loop sensorimotor integration for motor rehabilitation. Both virtual limbs and functional electrical stimulation (FES) were used as feedback, which could help patients improve the training performance through visual and sensory pathway. Results showed that five of ten patients obtained improvements of motor function. It provided evidences that the BCI-FES system is effective in restoring upper extremities motor function in stroke. In our future work, more cases are needed to demonstrate its superiority over conventional therapy and explore the potential role of MI in poststroke rehabilitation.

**Acknowledgments**

This work was supported in part by the Grant National Natural Science Foundation of China, under Grant Nos. 91 420302, 61573142. This work was also supported by the Programme of Introducing Talents of Discipline to Unive rsities (the 111 Project) under Grant B17017, the Fundamental Research Funds for the Central Universities (WH15 16018) and Shanghai Chenguang Program under Grant 14CG31.